\documentclass[3p]{elsarticle}
\usepackage{fleqn,amssymb,amsmath}
\usepackage{graphicx}
\usepackage{verbatim}
\newtheorem{thm}{Statement}
\def\stackunder#1#2{\mathrel{\mathop{#2}\limits_{#1}}}
\def\stackunder#1#2{\mathrel{\mathop{#2}\limits_{#1}}}
\newcommand{\beq}[2]{\begin{equation}\label{#1}
#2 \end{equation}}

\newcommand{\Req}[1]{(\ref{#1})}

\newcommand{\kvadrskob}[1]{\left[#1\right]}

\newcommand{\Lee}[1]{\stackunder{#1}{\rm L}}

\newcommand{\Const}{\mathop{\rm Const}\nolimits}

\newcommand{\xsort}[2]{\stackunder{#1}{#2}}

\begin{document}
\twocolumn[
\begin{frontmatter}
\title{The Possibility Of The Strict Global Thermodynamic Equilibrium In The Expanding Universe At Presence Of The Fundamental Scalar Field}
\author{Yurii Ignat'ev}
\address{Lobachevsky Institute of Mathematics and Mechanics, Kazan Federal University,\\ Kremleovskaya str., 35, Kazan 420008, Russia}


\begin{abstract}
In the article it is shown that at presence of fundamental scalar fields determining the masses of the scalar charged particles the global thermodynamic equilibrium (GTE) is compatible with a process of the cosmological expansion of the statistical system.
\end{abstract}

\end{frontmatter}
]
\section{The General Relativistic Theory And The Thermodynamic Equilibrium}

The basics of the general relativistic kinetic the\-ory (GRKT) were stated in 60's in the works of N.A. Chernikov \cite{chern1}, \cite{chern2} - \cite{chern5},
E.Tauber and J.Weinberg \cite{taub}, A.A.Vlasov \cite{vlas}, R.Lindqvist \cite{lindkv} and others.\\ Within the framework of the GRKT there were
formulated the general relativistic kinetic equations on which basis the macroscopic transport equations were built as well as the theory of
the global and local ther\-mo\-dy\-na\-mic equilibrium of the statistical system in gravitational and electromagnetic fields was de\-ve\-lo\-ped
\footnote{By the electromagnetic fields here one can imply also any vector fields.}. In the most complete and strict form the
results of these researches are explicated in the follo\-wing Author's works \cite{Yu_Pol,Yu_1982,Yukin1,Yukin2,Yukin3}. In this chapter we recap the results of these researches in the reduced form wherein we use more modern, canonical and explicitly invariant formulation of the kinetic theory, which was developed in Authors's works \cite{YuMono}.

\subsection{The Canonical Equations Of Motion In The Gravitational And Electromagnetic Fields
And The Macroscopic Fluxes}
The canonical equations of particle motion in canonically conjugate variables - coordinates $x^i$ and generalized momentum $P_i$ have form:
\begin{equation}\label{can_eq}
p^i\equiv\frac{dx^i}{ds}=\frac{\partial H_a}{\partial P_i};\quad \frac{dP_i}{ds}=-\frac{\partial H_a}{\partial x^i},
\end{equation}
where $H_a(x,P)$ is a relativistically invariant Ha\-milton function, $p^i$ is a kinematic momentum. %
The total derivative of an arbitrary function of the dy\-namic variables  %
$\Psi_a(x,P)$ with respect to the cano\-nical parameter $s$ is determined by the {\it Poisson brac\-kets}:
\begin{equation}\label{dpsi}
\frac{d\Psi_a}{ds}=[H_a,\Psi_a]\equiv \left[\frac{\partial H_a}{\partial P_i}\frac{\partial \Psi_a}{\partial x^i}- \frac{\partial H_a}{\partial x^i}\frac{\partial \Psi_a}{\partial P_i}\right].
\end{equation}
As a result of the equations (\ref{dpsi}) the Hamilton function itself is an integral of motion:
\begin{equation}\label{m2}
H_a(x,P)=\mathrm{Const}=\frac{1}{2}m_a^2,\Rightarrow (p,p)=m^2_a,
\end{equation}
where $m_a$ is a rest mass of particles. The relation (\ref{m2}) is called the momentum normalization ratio or
the equation of the mass surface\footnote{Hereinafter $(a,b)=g_{ik}a^ib^k$ is a scalar product of the momentums %
$a$ and $b$ relatively to the metrics $g_{ik}$.}.

For the charged particles with masses $m_a$ and electrical charges $e_a$ in a gravitational fields having metrics $g_{ik}(x)$
and electromagnetical fields with a vector potential  $A_i(x)$ %
the Hamilton function can be specified in the form\footnote{It is used everywhere the universal system of units $G=c=\hbar=k=1$ where $k$ is a Boltzmann constant.}:
\begin{eqnarray}\label{He}
H_a(x,P)=\frac{1}{2}g^{ik}(P_i-e_aA_i)(P_k-e_aA_k)\nonumber\\ \equiv \frac{1}{2}(P-e_aA,P-e_aA)\equiv \frac{1}{2}(P-e_aA)^2.
\end{eqnarray}

Let the statistical system in the gravitational field with metrics $g_ik$ and in the electromagnetic field with the potential $A_i$ %
consists of $N$ sorts of the identical particles and $f_a(x,p)$ are invariant functions of  distribution of these particles in the phase space $X\times P$, ($x^i$ --
are coordinates, $p^i$ is a momentum, %
so that:
\begin{equation}\label{n}
n^i_a(x)=\int\limits_{P(x)}f_a(x,p)p^idP_a
\end{equation}
there are $a$-sort particles' number flux densities
\begin{equation}\label{Na}
N_a(\tau)=\int\limits_V n^i_adV_i\equiv \int\limits_V dV_i \int\limits_{P(x)}f_a(x,p)p^idP_a
\end{equation}
and there is a number of $a$-sort particles on the spacelike hypersurface $V:\, dx^iu_i=0$, %
where
\begin{eqnarray}\label{dP}
dP_a=&{\displaystyle \frac{2S+1}{8\pi^3\sqrt{-g}}dp^1dp^2dp^3dp^4\delta(H_a)}\nonumber\\
\Rightarrow & dP_a={\displaystyle \frac{2S+1}{4\pi^3\sqrt{-g}}\frac{dp^1dp^2dp^3}{p_4}},
\end{eqnarray}
$H_a(x,P)$ is a Hamilton function of $a$-sort particle in the gravitational $g_{ik}$ and in the electromagnetic $A_i$ fields,
$S$ -- is a spin, $p_4$ is a positive root of the mass surface equation (the covariant time component of the momentum):
\begin{equation}\label{pp}
H_a(x,P)=\frac{1}{2}m^2_a\Rightarrow (p,p)=m^2_a.
\end{equation}

\subsection{The Common-relativistic Equations For Charged Particles}

Let further the following reactions run in the statistical system:
\begin{equation}\label{reactions}
\sum\limits_{A=1}^{m} \nu_A a_A \rightleftarrows
\sum\limits_{B=1}^{m'} \nu'_B a'_B \Rightarrow\sum\limits_k \sum\limits_{A=1}^{N} \nu^k_A a_A = 0\,,
\end{equation}
where $\nu_A$ is $a_A$-sort particle number of particles participating in the reaction; $A$ is a number of the particle, $||\nu^k_A||$ is an integer matrix of rank less than $N$ and the summation is carried out by all channels of reactions in which $a$-sort particles participate.

The distribution functions $f_a(x,p)$ are determined by means of the relativistic kinetic equations:
\begin{equation} \label{kin_eq}
[H_{a} ,f_{a} ]=I_{a} (x,P_{a} ),
\end{equation}
where $J_{a} (x,P_{a} )$ are the collision integrals:
\begin{eqnarray} \label{int_col}
I_{a} (x,P_{a} )=
-\sum\limits_a  \nu _{a} \int  '_{a} \delta ^{4}(P_{F} -P_{I} ) \nonumber\\
\times W_{IF} (Z_{IF} -Z_{FI} )\prod _{I,F} 'dP;
\end{eqnarray}
where
\[W_{FI} =(2\pi )^{4} 2^{-\sum  \nu _{A} +\sum  \nu '_{b} }|M_{IF} |^{2}  \]
is a scattering matrix for the channel of reactions \eqref{reactions}, ($|M_{IF}|$ are invariant amplitudes of scattering);
\[Z_{IF} =\prod _{I} f(P_{A}^{\alpha } )\prod _{F} [1\pm f(P_{B}^{\alpha '} )];\]
\[Z_{FI} =\prod _{I} [1\pm f(P_{A}^{\alpha } )]\prod _{F}
f(P_{B}^{\alpha '} ),\] sign ``+'' corresponds to bosons, ``-'' corresposnd to fermions (details see in \cite{Yukin1,Yukin2}).
\subsection{The Entropy Of The Statistical System}
The entropy of the statistical system on the hyper\-surface $V(\tau)$ is equal to:
\begin{eqnarray}\label{S}
S(\tau)=\sum\limits_a S_a(\tau) = \sum\limits_a \int\limits_{V(\tau)}u^idV_i \times\nonumber\\ \int\limits_{P_a(X)} [\pm (1 \pm f_a) \ln(1 \pm f_a) - f_a \ln f_a] dP_a,
\end{eqnarray}
where $\tau=\int u_idx^i$ is a proper time of an observer. The total derivative of the system entropy with respect to the time $\tau$
with an account of the optical theorem can be written in the form
\cite{YuMono}\footnote{for $T$ - invariant as well as for $T$ - nonvariant interactions.}:
\begin{eqnarray}\label{dS}
\frac{dS}{d\tau} = \sum \int \delta^{(4)}(P_F
- P_I)\times \nonumber\\
\ln\left(\frac{Z_{if}}{Z_{fi}}\right)(Z_{if} - Z_{fi}) W_{if}
\prod_{i,f} dP dV
\end{eqnarray}
Therefore for thermodynamical equilibrium $dS/d\tau=0$ a fulfillment of the functional Boltzmann equations is a necessary condition:
\begin{equation}\label{Bolts}
Z_{if}-Z_{fi}=0.
\end{equation}

\subsection{The Local Thermodynamic Equilibrium Of The System Of Electrically Charged Particles}

The next statements are true.
\begin{thm}
For the statistical system consisting of N sorts of identical particles to
be in the state of LTE in a gravitational field it is necessary and sufficient that
\begin{enumerate}
\item the invariant function of these particles' distri\-bution $f_a(x,p)$ has the locally equilibrium form:
\begin{equation}\label{loc_f}
f^0_a(x,p)=\left[\exp(-\gamma_a+(\xi,p))\pm 1\right]^{-1},
\end{equation}
where
\begin{equation}\label{xi}
\xi^i=\xi^i(x)\quad (\xi,\xi)>0
\end{equation}
is a timelike vector,
\item  the reduced chemical potentials $\gamma_a(x)$ satisfy the set of chemical equilibrium conditions
\begin{equation}\label{chem_eq}
\sum\limits_{A=1}^{N} \nu^k_A \gamma_A = 0\,,
\end{equation}
\end{enumerate}
\end{thm}
Locally equilibrium distribution functions (\ref{loc_f}) are the sole positively defined solutions of the Boltzmann functional equations (\ref{Bolts}). %
Introducing with a help of timelike vector $\xi^i(x)$ the macroscopic kinematic velocity \cite{sing}, $v^i$
\begin{equation}\label{ui}
v^i\equiv \frac{\xi^i}{\xi},\quad \xi\equiv \sqrt{(\xi,\xi)};\Rightarrow (v,v)=1,
\end{equation}
as well as the local temperature of the statistical system, $\theta$:
\begin{equation}\label{theta}
\theta(x)=\frac{1}{\xi(x)},
\end{equation}
and also introducing by means of the reduced che\-mical potential the che\-mical potential in the stan\-dard normalization:
\begin{equation}\label{mu}
\mu_a(x)=\gamma_a\theta,
\end{equation}
let us write in standard denotation the equilibrium distribution function and the
conditions of the che\-mical equilibrium:
\begin{equation}\label{f0}
f^0_a(x,p)=\left[\exp\left(\frac{-\mu_a+(v,P)}{\theta} \right)\pm 1\right]^{-1};
\end{equation}
\begin{equation}\label{chem_eq_mu}
\sum\limits_{A=1}^{N} \nu^k_A \mu_A = 0\,.
\end{equation}

\subsection{The Global Thermodynamic Equilibrium Of The System Of Electrically Charged Particles} \label{I.III.3}
The locally equilibrium distribution functions (\ref{loc_f}), (\ref{f0} turn the intergral of motion (\ref{int_col}) to the null equation.
In case when distribution functions (\ref{loc_f}) or (\ref{f0}) are the exact solutions of the kinetic equations (\ref{kin_eq})
the statistical system is in the strict {\it global thermodynamical equilibrium (GTE)}. At con\-di\-tions of GTE
the strict laws of conservation of each sort particles are fulfilled and the system en\-tropy is strictly constant $S = \Const$.
The substi\-tution of (\ref{loc_f}) to the kinetic equations
(\ref{kin_eq}) reduces them to the form:
\beq{3.2.1}{ \kvadrskob{H_a, \phi_a} = 0\,,
}
where
\begin{equation}\label{lin_int}
\phi_a(x,P) =\gamma_a(x)+(\xi,P_a)
\end{equation}
is an argument of the equilibrium distribution fun\-c\-tion (\ref{loc_f}).
Thus to ensure GTE there should exist a linear integral of motion having $\xi^i$ is a timelike vector.
Resolving \Req{3.2.1} we obtain a set of necessary and sufficient conditions of GTE existence GTE \cite{taub}, \cite{chern4,chern5}, \cite{Yu_Pol,Yukin3}:
\begin{eqnarray}
\label{Leeg}
\Lee{\xi} g_{ik}& =& %
\left\{\begin{array}{ll}
0\,, & m_a\not=0;\\
\varrho g_{ik}\,, & m_a=0,
\end{array}\right.
  \\
\label{LeeA}
e_a \Lee{\xi} A_i & =& \gamma\,_{a,i}\,,
\end{eqnarray}
where $\Lee{\xi}$ is a Lie derivative with respect to the direc\-tion $\xi$, $\rho(x)$ -- is an arbitrary scalar function.

Let us notice a very notable for the cosmology strict fact that follows from (\ref{Leeg}) and have a power of theorem.

\begin{thm}\label{stat1}If the rest masses of all the particles of the statistical system are equal to zero then the condition of the global thermodynamic equilibrium is possible if and only if the metrics is conformally stationary. If even a single sort of massive particles is present in the statistical system then the condition of the GTE of the statistical system is possible only for the stationary gravitational field.\end{thm}

Thus if in the statistical system with electro\-magnetic interaction there is at least single sort of the particles with a non-zero rest mass then this particle's motion vector $\xi^i$ should be {\it a timelike\\ Killing vector} (\ref{Leeg}).

Next, for the cosmology it is of extremely great importance the property of the asymptotic conformal invariance of the relativistic theory of the statistical systems
with electromagnetic interaction which was proved in the Author's work \cite{Yu_1982}:
\begin{thm}\label{stat2}
In the ultrarelativistic limit
\begin{equation}\label{ultra}
m_a/\langle p_a\rangle\to0
\end{equation}
the kinetic theory is asymptotically conformally in\-va\-riant i.e.
the kinetic equations with accuracy within values of the second
order of smallness of parameter (\ref{ultra}) are invariant with respect to the conformal tran\-sformations %
 \begin{eqnarray}\label{confg}
 d\bar{s}^2=\rho^2ds^2;\\
 \label{confA}
 \bar{A}_i=A_i+\partial_i\varphi;\\
 \label{confP}
 \bar{P}_i=P_i-e\partial_i\varphi.
 \end{eqnarray}
\end{thm}

As a result of the asymptotic conformal inva\-riance of the relativistic kinetic theory the state of GTE is asymptotically reachable in the ultrare\-la\-tivistic cosmological plasma. The violation of this state in consequence of plasma cooling and violation of the condition (\ref{ultra}) triggers the me\-cha\-nism of spon\-taneous breaking of different symmetries in the Uni\-verse including the barion symmetry. Such me\-cha\-nisms were investigated by several authors in 1970 and 1980 \cite{Weinberg}, \cite{Fry1} and others and more strict kinetic theory of the barion symmetry spontaneous breaking was developed in the Author's papers \cite{Yu_Bar85,Yu_Bar05_1,Yu_Bar05_2,Yu_Bar07}.

\subsection{The Equilibrium Self-Gravitating Statistical\\ Systems}
Since all moments of the equilibrium distribution function are determined via scalars $\xi^2$,
$\stackunder{a}{\lambda}$, $\Phi$ and tensors $\xi^i$, $g^{ik}$,
$\xi^i \xi^k$, ..., then the conservation laws for the moments of the distribution function are also fulfilled \cite{Yukin3}:
\beq{Leen}{ \Lee{\xi} n^i_a =
0\,; }
\beq{LeeT}{\Lee{\xi} \stackunder{a}{T}^{ik} = 0 }
etc.

In consequence of the first group of equations \Req{Leeg} the following components of Riemann-Ricci\\ tensor and Einstein tensor are conserved along the direction
$\xi^i$:
\beq{LeeR}{\Lee{\xi} R_{ijkl} = 0\,; \quad \Lee{\xi}
R_{ij} = 0\,; \quad \Lee{\xi} G_{ij} = 0\,.}
Therefore in consequence of the Einstein equations there should be fullfilled the next relations:
\beq{LeeEinst}{\Lee{\xi} T_{ij} = 0\,.}
This allows us to carry out the strict
classification of all the equilibrium self-gravitating
configurations with electromagnetical interaction
 \cite{Yu_GTE}:
\begin{thm}
Let $\xi^i(x)$ is a timelike Killing vector in the equilibrium distribution function
 (\ref{loc_f}) and the gravitational field of the equilibrium configuration allows
 group of motions of the  $r$, $G^r$ order so that:
\begin{equation}\label{LeeGr}
\Lee{\lambda}g_{ik}=0; \quad \lambda=\overline{1,r},
\end{equation}
where $\Lee{\lambda}\equiv \Lee{\xi_\lambda}.$
Let us expand the Killing vector by vectors of $G^r$ group:
\begin{equation}\label{xi_a}
\xi^i = \sum\limits_{\lambda=1}^{r} \alpha_{\lambda}
\xsort{\lambda} {\xi}^i\,; \quad (\alpha_{\lambda} = \mathrm{Const})\,.
\end{equation}
Then for the global equilibrium of the self-gravitating statistical systems of particles with electromagentic interaction it is necessary and
sufficient that the Killing motion vector of the system $\xi^i$ and tensor of the electromagnetic field $F_{ik}$ are conserved during motions along the group $G^r$:
\begin{eqnarray}\label{centre}
\Lee{\lambda}\xi^i =0\Rightarrow \nabla_k\sum\limits_{\lambda,\lambda'\not=}\xsort{[\lambda} {\xi}^k\xsort{\lambda']} {\xi}^i;\\
\label{Fxi}
\Lee{\lambda}F_{ik}=0; \quad \lambda,\lambda'=\overline{1,r}.
\end{eqnarray}
\end{thm}
In the language of generators of Lie group $X_\lambda=\xsort{\lambda} {\xi}^i\partial_i$ the condition (\ref{centre}) means that the group of motion of the charged particles' equilibrium self-gravitating configuration should have a timelike center \cite{IvanovGTE}.

\section{The Global Thermodynamic Equilibrium Of The Statistical Systems With Scalar Interaction}\label{I.III.5}
The common-relativistic kinetic theory of the statistical systems with scalar interaction was formu\-lated by the Author and above cited papers \cite{Yu_1982,Yukin1,Yukin2,Yukin3} and in more modern form taking into account fantom scalar fields this theory was refor\-mulated in \cite{YuScal12_1} -- \cite{Ignat14_2}. In particular in these papers it is shown that the invariant Hamilton function of the particle in a scalar field
 $\Phi$ can be chosen in the next form accurate within automorphisms:
\begin{equation}\label{HF}
H(x,P)=\frac{1}{2} \left[(m^*)^{-1}(\Phi)(P,P)-m^*(\Phi) \right]=0,
\end{equation}
where
\begin{equation}\label{m^*}
m^*(\Phi)=|m+q\Phi|\equiv |\phi|,
\end{equation}
$m$ is a bare mass of a particle, $q$ is particle scalar charge.
Function $m^*(\Phi)$ can be named as particle effective mass. Thus the normalization ration for the generalized momentum is fulfilled:
\begin{equation}\label{PPm}
(P,P)={m^*}\ \!^2.
\end{equation}
The kinetic equations (\ref{kin_eq}) at that conserve their form. The relations for the entropy (\ref{S}) and its derivative (\ref{dS}) also
conserve their form thereby the formula for the locally equilibrium distribution fun\-c\-tion is also conserved (\ref{loc_f}). %
The substitution of this function in the kinetic equations (\ref{kin_eq}) with an account of the explicit form of the Hamilton function (\ref{HF})
leads to the next relation:
\[
-\gamma_{a,i}\frac{P^i}{m^*_a}+\frac{1}{m^*_a}\xi_{i,k}P^iP^k-\mathrm{sgn}(\phi_a)q_a\xi^i\Phi_{,i}=0,
\]
for the fulfillment of which as a consequence of the generalized momentum the necessary and sufficient conditions are:
\begin{eqnarray}\label{gammaconst}
\partial_i\gamma_a=0\Rightarrow \gamma_a=\mathrm{Const};\\
\label{conf}
\Lee{\xi}g_{ik}=\sigma g_{ik};\Rightarrow\\
\label{sigma}
\sigma=\Lee{\xi}\ln m^*_a.
\end{eqnarray}
The substitution of (\ref{sigma}) in (\ref{conf}) leads to the condition:
\begin{equation}\label{gconfa}
\Lee{\xi}\frac{1}{m^*_a}g_{ik}=0.
\end{equation}

In the case if particle bare masses $m\equiv m_a$ in (\ref{m^*}) are different from zero
there remains the single possibility of
resolution of the relations (\ref{conf}) -- (\ref{sigma}) as a consequence of the difference of the bare masses and particle charges:
\begin{eqnarray}\label{leeg}
\Lee{\xi}g_{ik}=0;\\
\label{flee}
\Lee{\xi}\Phi=0.
\end{eqnarray}
Exactly this conclusion was made even in former works of the Author like \cite{Yukin3} and others. %
Thus also in the case of scalar interactions for the maintenance of the thermodynamic
equation it was required the stationarity of the gravitational and scalar fields. However this conclusion is not absolutely strict one. %
In case if bare masses of all particles are equal to zero there exists one more possibility of the global thermodynamic
equilibrium since in this case it is
 $m^*_a=|q_a\Phi|$:
\begin{equation}
\Lee{\xi}\frac{1}{\Phi}g_{ik}=0.
\end{equation}
Apparently Author halted with investigation of this case in 1982-1983 due to extreme exoticism
of the suggestion about all bare masses of elementary par\-ticles could be equal to zero. However in 2014 after the discovery of the Higgs boson
such a suggestion is now natural. Thus let us formulate
the statement about the global thermodynamic equilibrium
of the statistical system with scalar interaction.
\begin{thm}
If the effective masses of all particles of the statistical systems are generated exclusively by the scalar interaction then for the global ther\-mo\-dynamic equilibrium
of such system it is necessary that the conformal metrics of the space-time
\begin{equation}\label{g_}
\bar{g}_{ik}\equiv \frac{1}{|\Phi|}g_{ik}
\end{equation}
is a stationary one and chemical potentials of the statistical system are constant.
\end{thm}

This property of the statistical systems with a scalar interaction can bring
the far-reaching conse\-quences for the cosmology. Let us notice that if consider for instance the cosmological models
with a constant curvature:
\begin{equation}\label{cosm}
ds^2=a^2(\eta)(d\eta^2-dl^2),
\end{equation}
where $dl^2$ is a metrics of the 3d space of the constant curvature then
the condition (\ref{g_}) brings us to the following relation for the potential of the scalar field:
\begin{equation}\label{Fa}
|\Phi|\sim a^2(t).
\end{equation}
In such system the strict global thermodynamic equilibrium will always be maintained. Let us notice that
the presence of vector interactions can not change this conclusion and this one can find in the above cited Author's works.
The presence of such a stage in the evolution of the Universe can lead to the restoration of the barion and other symmetries at this stage.


\end{document}